\documentclass[twocolumn,showpacs,preprintnumbers,amsmath,amssymb,aps,prl]{revtex4}
\usepackage{graphicx}

\begin{document}

\title{Dynamic Phases, Pinning, and Pattern Formation  for Driven Dislocation Assemblies}
\author{C. Zhou, C. Reichhardt, C.J. Olson Reichhardt, and I.J. Beyerlein}
\affiliation{Center for Nonlinear Studies and Theoretical Division, Los Alamos National Laboratory, Los Alamos, New Mexico 87545, USA}

\date{\today}
\begin{abstract}
We show that driven dislocation assemblies 
exhibit a set of dynamical phases remarkably 
similar to those of driven systems with quenched disorder
such as vortices in 
superconductors, magnetic domain walls, and charge density wave materials.
These phases include jammed, fluctuating, and 
dynamically ordered states, and each produces
distinct dislocation patterns as well as 
specific features in the noise fluctuations and transport properties.
Our work suggests that many of the results established for
systems with quenched disorder undergoing depinning transitions can be
applied to dislocation systems, providing a new approach for understanding
dislocation pattern formation and dynamics.
\end{abstract}
\maketitle

\vskip2pc
There are numerous examples of systems of collectively interacting particles
that, when driven externally, 
depin and undergo dynamical pattern formation and/or
dynamic phase transitions, such as a transition from a fluctuating to a
nonfluctuating state.
Such 
systems include
domain walls, driven vortices in type-II superconductors
\cite{3bhattacharya,4koshelev,5olson,8pardo,9hellerqvist,25troyanovski},
sliding charge density waves \cite{7danneau},
and driven Wigner crystals \cite{EC}.
In these systems, fluctuating and intermittent dynamics arise just above
depinning when an applied external force is increased from zero,
while for higher drives the particles dynamically order into patterns such
as anisotropic crystals or moving smectic phases with different types
of fluctuation statistics \cite{26yaron,6faleski,23balents,24ledoussal,18moon}.
Dislocations in materials are known to undergo a transition at the onset of
irreversibility or yielding that has similarities to depinning \cite{laurson2010,moretti2004};
however, it has not been shown whether driven dislocations can exhibit
the same general features as other systems with depinning transitions.
Establishing such a connection could potentially open
an entirely new paradigm for understanding driven dislocations. 

It is known that organized dislocation structures
within individual crystals, such as tangles, cells, or planar walls,
can become more refined and better defined as stress or
strain increases.  Two-dimensional (2D) and
three-dimensional (3D) dislocation dynamics simulations
based on linear elasticity theory predict self-organization of
dislocation assemblies into varying configurations,
such as pileups near the yielding or depinning transition
\cite{moretti2004,laurson2010,bako2008}
and 2D mobile walls
\cite{2miguel,12miguel}
or 3D slip bands
\cite{csikor2007,2008wang}
under an
external drive.  
Below a critical stress where dislocations show no net motion,
the system is considered jammed \cite{1tsekenis,1tsekenisA}, while 
intermittent
or strongly fluctuating behavior with highly jerky or avalanche-like
motion occurs
above the
critical stress
\cite{2miguel}. 
Avalanche behavior
with power-law velocity distributions 
is proposed to be a signature of
critical dynamics
\cite{2miguel,12miguel,16dimiduk,14}.
No correlations between the transitions in patterning
and the intensity of applied stress or strain have been
established
before now.
Here we demonstrate that driven dislocation assemblies exhibit the
same nonequilibrium phases observed for collectively interacting particle
systems exhibiting depinning,
including pattern organization in the  pinned state,
a strongly fluctuating intermittent phase with a coexistence of
pinned and moving particles
\cite{3bhattacharya,5olson,6faleski,20reichhardt,21field},
and at higher drive, when the effectiveness of the substrate is reduced,
a phase in which the dislocations organize into moving
wall structures
\cite{4koshelev,5olson,8pardo,25troyanovski,7danneau,23balents,24ledoussal,18moon}.
The onsets of these different dynamical regimes are
correlated with pronounced changes in the transport curves
\cite{3bhattacharya,9hellerqvist},
noise properties \cite{5olson,10marley}, and
spatial structures \cite{8pardo,25troyanovski,26yaron}.
The onset of these phases can be observed via changes in
the dislocation structure, mobility, velocity distribution, and velocity noise.
Our work implies that many of the established results
obtained for driven vortex and other systems
can be used to understand dislocation dynamics.

We utilize a discrete dislocation dynamics 
model with periodic boundary conditions
for a
2D cross section 
of a sample
containing $N_D=480$ 
straight edge dislocations that glide along parallel slip planes.
This model was previously shown to capture
the behavior observed in stressed anisotropic materials,
particularly the intermittent flow near the onset of
motion \cite{1tsekenis,1tsekenisA,2miguel}.
An equal number of positive and negative moving dislocations
are randomly
placed in the sample and can move in the positive or negative
$x$-direction depending on the
sign of their Burgers vector ${\bf b}$.
At most one dislocation is allowed to reside on a plane, so in-plane
pile-ups are prevented.  The dislocations are also restricted from leaving
their assigned glide plane. 
Rather than imposing an annihilation rule 
\cite{laurson2010,2miguel,12miguel},
we enforce that two adjacent glide planes
must be separated by
at least $\delta y$, where $\delta y$ is on the order of the
Burgers vector of the dislocations
\cite{1tsekenis,1tsekenisA,2011ispanovity}.

The dislocations
interact via a long-range anisotropic stress field that is
attractive between two oppositely signed dislocations and
repulsive for liked-signed pairs.  We utilize a replicated
image model to allow a large number of dislocations to be
simulated efficiently over long times
\cite{hirth}.
Within the simulation volume, all dislocations are subject to the stress
fields of all surrounding dislocations regardless of their position.
To best make the connection with particle systems, nucleation of
fresh dislocations during loading is suppressed.
Under an external
applied stress $\tau_{\rm ext}$, dislocation $i$ moves along $x$ in its
assigned plane according to an overdamped equation of motion given by
$\eta\frac{dx_i}{dt}=b_i\left(\sum_{j\neq i}^N \tau_{\rm int}({\bf r}_j-{\bf r}_i)-\tau_{\rm ext}\right)$
where $x_i$ is the $x$ coordinate of dislocation $i$ at point
${\bf r}_i=(x_i,y_i)$ with Burgers vector value $b_i$, $\eta$ is the effective
friction, and $\tau_{\rm int}({\bf r}_j-{\bf r}_i)$ is the long-range shear
stress on dislocation $i$ generated by dislocation $j$.
The external load
on a dislocation is proportional to the stress, $F_{d} = b\tau_{ext}$.
For ${\bf r}=(x,y)=(x_j,y_j)-(x_i,y_i)$, $\tau_{\rm int}({\bf r}_j-{\bf r}_i)$
for an edge dislocation with Burgers vector value $b$ is
$\tau_{\rm int}({\bf r})=b\mu[x(x^2-y^2)]/[2\pi(1-\nu)(x^2+y^2)^2]$
where $\mu$ is the shear modulus and $\nu$ is the Poisson's ratio.  The
length of the square simulation cell $L$ is set to unity and the simulation
volume remains fixed throughout loading.  We normalize 
our units such that $b=1$, $\eta=1$, and $\mu/2\pi(1-\nu)=1$.  
The system is initially allowed to relax without 
an applied external drive. 
After relaxation, 
the external drive is applied quasi-statically, with sufficiently long waiting times between increments to avoid transient effects. 
We measure the average absolute value of the dislocation velocities
$\langle |v|\rangle$ as a function of the stress.
This is analogous to the voltage versus applied current curve
for vortices in superconductors.

In order to detect and
characterize the dislocation content and charge of the wall structures, we
examine the distribution of $d_x=|x_i-x_j|$, the $x$-axis separation between
two dislocations lying on adjacent planes.  The fraction of dislocation pairs
with $d_x<w$ is denoted as $P^w=n_{d_x<w}/n_{tot}$, where $w$ is the pre-assigned
maximum wall width, $n_{tot}$ is the total number of pairs in this system,
and $n_{d_x<w}$ is the number of pairs satisfying $d_x<w$.  
Here
we set $w=0.05$, although other reasonable values, such as $w=0.02$, give
qualitatively similar results.  To distinguish unipolar from dipolar walls,
we 
discriminate between those pairs of like and
unlike sign that lie within the critical wall width.  The difference
$B =P_{++,--}-P_{+-}$ between the fraction of pairs of like 
(unlike) sign $P_{++,--}$ ($P_{+-}$)
is directly related to the
net Burgers vector around one dislocation within the wall width $w$.  Thus,
when a dipolar wall forms, $B$ approaches zero since the number of
unlike sign pairs is almost the same as the number of like sign pairs.
When a unipolar wall forms, $P_{+-}$ is zero and 
$B$ is equal
to $P_{++,--}$.

\begin{figure}
\includegraphics[width = 3.5in]{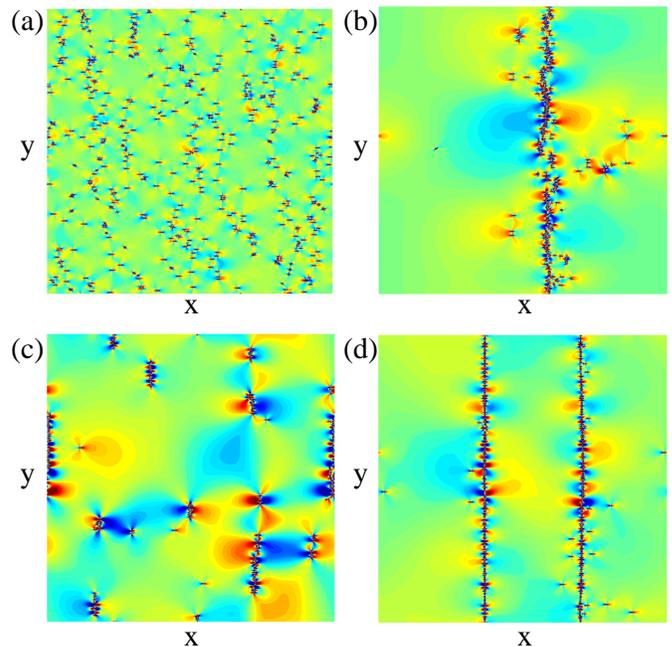}
\caption{
Stress maps of the sample range from large
negative (blue) to large positive (red) stress. 
(a) The initial dislocation positions at zero load.
(b) Just before yielding, the dislocations
are predominantly located at pile-ups to form a single bipolar wall.
(c)
Above yielding at $F_d=3.6$, the wall breaks apart and the structure
exhibits intermittent dynamics.
(d) At $F_{d} = 8.0$ there is a dynamical ordering
into polarized walls,
each composed of dislocations with the same Burgers vector orientation.
}
\end{figure}

As the system of randomly positioned dislocations is allowed to relax
under zero applied stress, the dislocations reassemble into a locked
configuration determined by the long-range stress fields
they collectively produce.  The relaxed arrangement of the
dislocations shown in Figure~1(a) is disordered
and contains no percolating walls.
The internal stresses generated by this
spatially random arrangement are high and are distributed uniformly
across the volume.

For loads $0 < F_{d} < 2.0$, the
dislocation pattern slowly changes after each load increment
but 
$\langle |v|\rangle$ goes to zero in the long time
limit,
indicating that the system is in the jammed phase below
the critical yield \cite{1tsekenis,1tsekenisA}. 
Under these low drives, any dislocation motion merely causes the dislocations to
lock into another immobilized pattern. 
Figure~1(b) illustrates a typical locked dislocation configuration for loads
just below critical yield (i.e., $F_{d} < F_{c}$), where most dislocations have assembled into a
dipolar wall comprised of a disordered arrangement of positive and
negative dislocations that cannot move past one another.
Compared to the initial state [Fig.~1(a)],
the internal stress
[Fig.~1(b)] remains high but is more localized,
and large stress concentrations appear in the vicinity of the wall.
Such walls are analogous to the model of a ``polarized'' wall
\cite{1987mughrabi,1996kuhlman},
with dislocations of predominantly one sign on one side of the wall
and the other sign on the other side.
They are thought to be responsible for the observed hysteresis in
unloading or the Bauschinger effect in subsequent reverse
loadings
\cite{1980kocks,1990stout}.
Observations of polarized walls have been reported in crystals
deformed to large strains
\cite{2007xue,2001huang,2000hughes}.

Just above yielding, the dipolar wall structure breaks down
as shown in Fig.~1(c) and the system enters a state
characterized by strong fluctuations in the dislocation positions.
The dipolar walls repeatedly break up and reform,
while the remaining wall fragments
become 
smaller at higher drives
and show continual change. 
The fluctuating state persists up to
$F_{d} = 5.0$,
when a new type of dynamic pattern appears where the dislocations
form unipolar walls composed of only one type of
dislocation, either negative or positive, as shown in Fig.~1(d).
These walls can be identified as disordered tilt walls.
An ideal model of a tilt wall involves a periodic array of edge
dislocations that accommodate a tilt misorientation between two
adjoining crystals. The development of low-misoriented tilt
walls is suspected to be a precursor for the eventual formation of
subgrains in heavily deformed crystals
\cite{2004dallatorre,2006dallatorre}.
Most importantly, when the unipolar walls form, the internal stress
decreases in extent and intensity.  The alternating positive and negative
stress pattern that develops along the wall in Fig.~1(d) is consistent
with the theoretical prediction for an infinite array of perfectly
aligned, like-signed edge dislocations
\cite{Libook}
from linear elasticity theory.  Thus, the high applied drive enables
the dislocations to assemble into a low energy ordered structure,
a result that is consistent with the theories of substructure
development proposed by Kuhlman-Wilsdorf
\cite{1996kuhlman}.

\begin{figure}
\includegraphics[width = 3.5in]{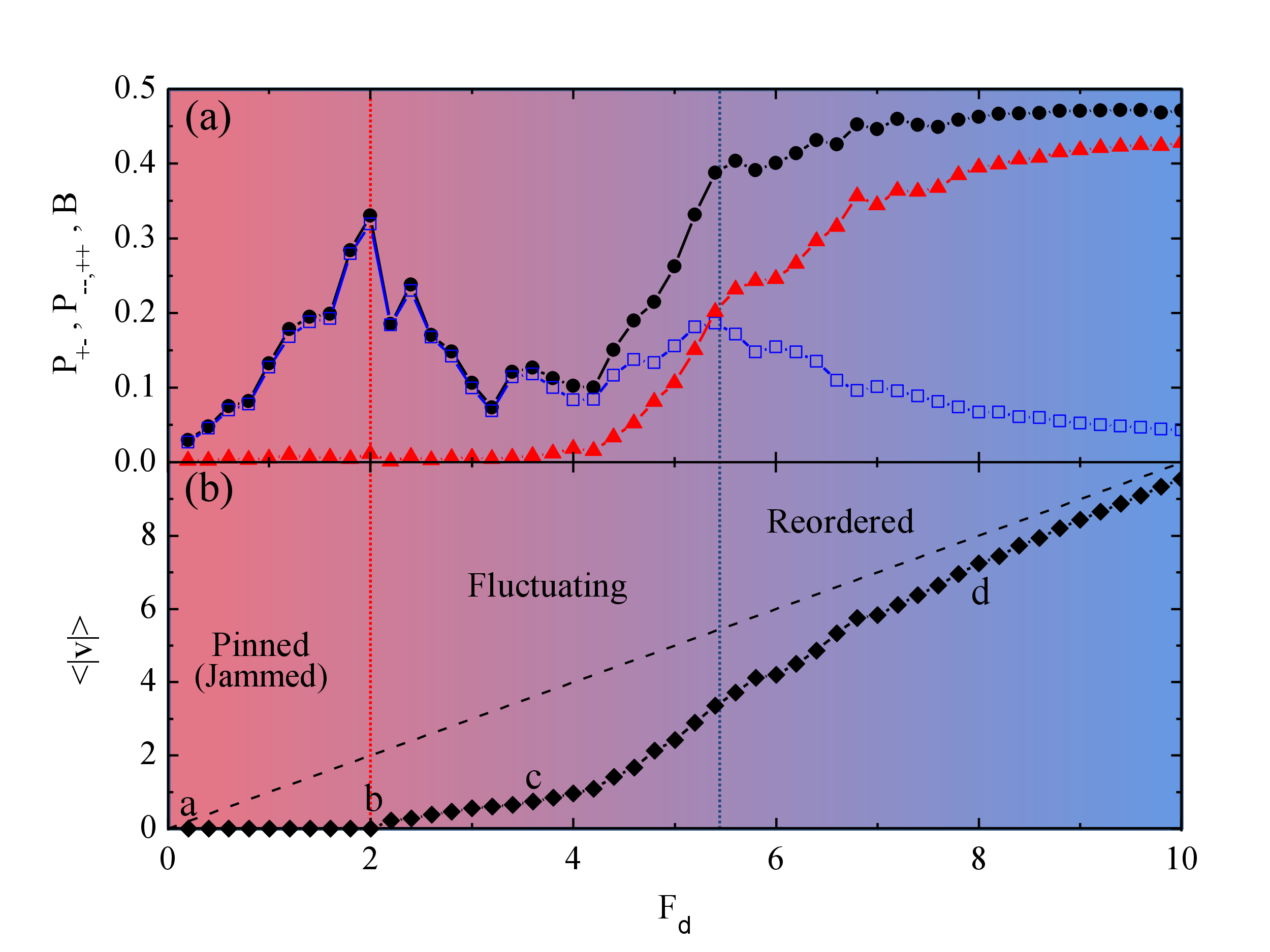}
\caption{
(a)
$P_{+-}$ (blue squares),
the fraction of dipolar walls,
vs $F_d$
has a peak just below yielding. 
$P_{--,++}$ (black circles),
the fraction of 
uni-polar walls, 
passes
through a plateau when the polarized wall state forms.
$B$ (red triangles) is a measure of the net Burgers vector in the
walls.  
(b)
The average absolute value of the dislocation velocity $\langle |v|\rangle$
(solid lower curve) vs $F_{d}$. The upper dashed
curve shows 
$\langle |v|\rangle$ for non-interacting dislocations.
Visible in the lower curve is a
yielding point, a nonlinear region corresponding to the disordered or
fluctuating
regime, and a linear region at high drives when the system is
dynamically ordered.
Points a, b, c, and d indicate the $F_{d}$ values illustrated in Fig.~1.
}
\end{figure}

In Fig.~2 we show that the changes in the dislocation structure
produce signatures in
$\langle |v|\rangle$
versus 
$F_{d}$ 
for the system in Fig.~1.  The upper curve in Fig.~2(b) shows the
simple linear dependence
of $\langle |v|\rangle$ on $F_{d}$ expected
for a single dislocation. 
For
the interacting system, $\langle |v|\rangle$ is zero below yielding for
$0.0 < F_{d} < 2.0$, increases nonlinearly
for $2.0 \leq F_{d} < 7.5$, and then starts to become linear again
for $F_d \geq 7.5$.
Figure~2(a) characterizes the ordering dynamics as
a function of $F_{d}$. 
Since the system is initialized in a random state containing no walls,
$P_{+-} \approx 0$ 
at $F_d=0$,
but as the load increases, $P_{+ -}$
reaches a maximum just below the yielding point
as shown in Fig.~1(b) where a large dipolar
wall forms.  Above yielding, $P_{+ -}$ decreases
in the fluctuating regime when the walls break up,
and it gradually drops to zero in the high-driving region where the
unipolar walls form.
To identify the formation of the unipolar dislocation walls, we measure 
$P_{++,--}$, which rises
for $F_{d} > 5.0$
in Fig.~2(a).
Also shown in Fig.~2(a) is the net Burgers vector of the walls,
indicating that for $F_{d} > 5.0$ the walls are
indeed unipolar and contain either exclusively
positive or negative dislocations.
Samples with smaller numbers of dislocations show the same
general features in
$P_{+-}$, $P_{++,--}$, and their difference $B$, 
as shown in Suppl.~Fig.~1 \cite{Suppl}.

The overall dynamics illustrated in Fig.~1 and Fig.~2 are remarkably
similar to those
observed in driven systems with quenched disorder.  For example, for
vortex matter as a function of external drive, there
is a low drive pinned phase, a strongly fluctuating phase where the
vortex lattice structure is disordered, 
and
a highly driven phase where
dynamical pattern formation occurs \cite{4koshelev,5olson,18moon}.
The corresponding vortex velocity-force curves also show
the same features: the fluctuating
phase is correlated with a nonlinear region, while
in the dynamically reordered phase the velocity depends
linearly on the drive \cite{3bhattacharya,5olson,6faleski,9hellerqvist}.

\begin{figure}
\includegraphics[width = 3.5in]{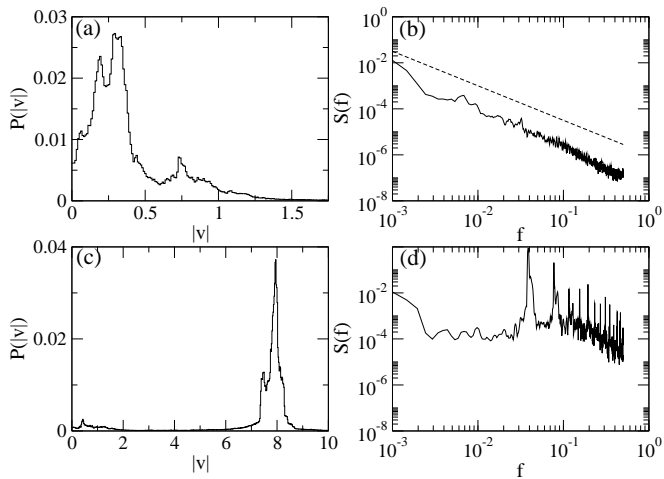}
\caption{
(a) 
$P(|v|)$
at $F_{d} = 3.2$ is bimodal in the fluctuating phase.
(b) The corresponding 
$S(f)$ from the
time series of the velocity has a $1/f^{1.5}$ shape.
(c)
$P(|v|)$ for $F_{d} = 8.0$ in the ordered phase shows
a sharp peak.
(d) The corresponding 
$S(f)$ has
a characteristic peak indicating narrow band noise.
}
\end{figure}

The dynamical phases in the vortex system have also been characterized
by changes in the velocity noise fluctuations
across different regimes. Just above depinning in the fluctuating regime,
there is a strong $1/f^\alpha$ noise signal \cite{3bhattacharya,5olson}
associated with a bimodal velocity distribution that indicates a coexistence
of pinned and moving vortices \cite{5olson,6faleski,17pertsinidis}.
The onset of dynamical ordering is accompanied by a
drop in the noise power $S_{0}$ as well as the appearance of narrow band
noise features \cite{5olson,10marley}.
For the dislocation system,
in Fig.~3(a) we plot the instantaneous velocity distribution
$P(|v|)$
in the fluctuating phase at $F_{d} = 3.2$.  We find a bimodal velocity
distribution that appears because a portion of the dislocations are immobilized
in pileups while other dislocations have broken out of pileups and
are mobile.
A similar bimodal velocity distribution appears
in the fluctuating phase for driven vortices and colloids.
The corresponding
power spectrum $S(f)$ presented in Fig.~3(b) of the time
series of the average dislocation velocity has a
$1/f^{1.5}$ signal in this regime,
in good agreement with studies of driven vortices.
At $F_{d} = 8.0$ [Fig.~3(c)] in the dynamically ordered phase,
$P(|v|)$
has a single sharp peak and
the corresponding $S(f)$
in Fig.~3(d) has a narrow band feature
with a characteristic frequency generated by the formation of an
ordered structure of unipolar walls.
In Fig.~4(a) the noise power $S_{0}$ for a fixed frequency
reaches a peak in the middle of the fluctuating disordered phase and then
decreases as the dynamically ordered phase of unipolar walls is approached.

By conducting a series of simulations for varied dislocation densities, $\rho = N_D/L^2 $, and
analyzing the ordering dynamics, we construct the dynamic phase diagram
shown in Fig.~4(b). The lower curve indicates the
yielding transition from the low drive
jammed or pinned phase of dipolar walls to the
fluctuating disordered phase.
The onset of the dynamically ordered phase is defined
as the force at which the unipolar wall structures start to form, and is
plotted in the upper curve.
As 
$\rho$ increases, the yielding point rises to higher
$F_d$
since the dislocations have a more difficult time breaking through the
dipolar walls that form.
The increase in yield threshold with increasing $\rho$ remains robust when we
perform simulations with different initial dislocation configurations.
Fig.~4(b) shows that the onset of the high drive dynamically ordered phase
also increases in a similar fashion with increasing $\rho$.
This phase diagram exhibits the same features
observed for vortex systems as a function of pinning strength vs
external drive, where both the critical depinning force and the onset of the
ordering rise to higher drives with increasing pinning strength
\cite{18moon}.
For the dislocation system, increasing
$\rho$ is equivalent to increasing the pinning strength. 
In our simulation, 
the pinning strength is directly related 
to the dislocation dipole break stress given by
$\mu$/(8$\pi$dy(1-$\nu$)) \cite{1960li}, 
where dy $\propto 1/\rho$. 
Thus, as Fig.~4(b) shows, the pinning strength 
scales as $\rho$.  This differs
from the recent simulation results obtained 
in smaller systems \cite{1tsekenis,1tsekenisA}. 

\begin{figure}
\includegraphics*[width=3.5in]{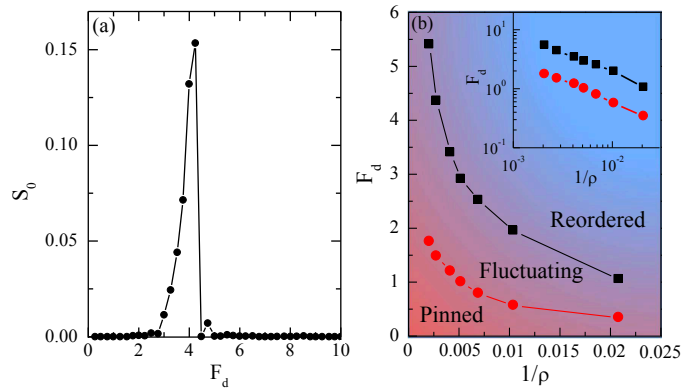}
\caption{
(a)  
$S_0$ vs $F_d$ at a fixed frequency of 
$f  = 5 \times 10^{-3}$ 
has
a peak 
in the fluctuating phase
and drops as the dynamically ordered phase is approached.
(b)
The dynamical phase diagram $F_{d}$ vs $1/\rho$,
where $\rho$ is the dislocation density.
The lower curve (red circles) indicates the onset of yielding and the upper
curve (black squares) is the onset of the dynamically induced ordered phase;
the fluctuating phase falls between the two curves.
Both the critical yielding and the dynamical ordering shift to
higher drives as the 
$\rho$ increases.
Inset: The same curves plotted on a log-log scale.}
\end{figure}

The type of ordered state that forms in the strong driving
regime varies depending on the rate at which the external load is applied. The
ordered polarized walls illustrated in Fig.~1
form under continuous sweeps of the load.  If the load is
instead instantaneously set to a high value, we observe a transient
disordered phase followed by the formation of
multiple lower density unipolar walls,
as illustrated in Suppl.~Fig.~2 \cite{Suppl}, instead of 
the two unipolar walls shown in Fig.~1(d).
As for the nature of the yielding transition, recent work on depinning
systems has suggested that 
plastic depinning falls into the class of absorbing phase transitions,
specifically directed percolation \cite{20reichhardt,27okuma},
suggesting that the onset of yielding for dislocation systems
could also fall into the class of directed percolation.

In summary, we have shown that driven dislocation assemblies
exhibit the same nonequilibrium phases observed
for systems of collectively interacting particles such as
vortices in disordered superconductors.
These include a jammed phase analogous to a pinned state,
a fluctuating or disordered
phase, and dynamically ordered or pattern forming states.
All of the states are associated with transport signatures such as
changes in the transport noise fluctuations as
well as features in the dislocation velocity vs applied shear, in analogy
with velocity-force curves.

We acknowledge helpful discussions with Karen Dahmen.
This work was carried out under the auspices of the NNSA of the US DoE at
LANL under Contract No. DE-AC52-06NA25396.

\vfill\eject


\section{Supplementary Information for ``Dynamic Phases, Pinning,
and Pattern Formation for Driven
Dislocation Assemblies''}
\subsection{C. Zhou, C. Reichhardt, C.J. Olson Reichhardt, and I.J. Beyerlein}

\setcounter{figure}{0}

\begin{figure}
\includegraphics[width = 3.5in]{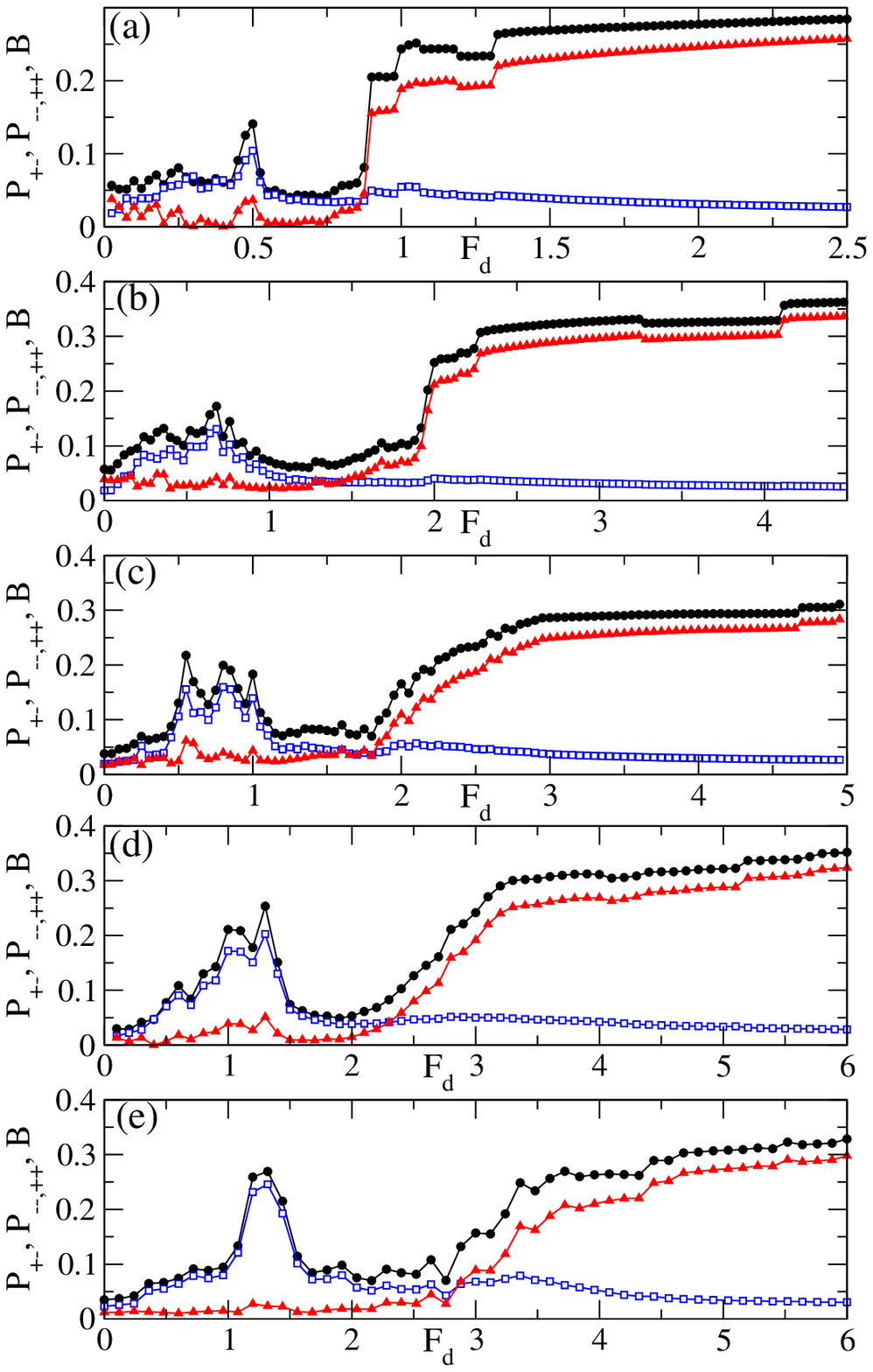}
\caption{(Supplemental)
$P_{+-}$ (blue squares), the fraction of walls containing dislocations with both
positive and negative Burgers vectors, 
$P_{--,++}$ (black circles), the fraction of walls composed of a single polarity
of burgers vectors, 
and $B$, the net Burgers vector in the walls,
vs $F_d$ for samples with different numbers of dislocations $N_D$.
{\bf a} $N_D=48$.
{\bf b} $N_D=96$. 
{\bf c} $N_D=144$.
{\bf d} $N_D=192$.
{\bf e} $N_D=240$.
{\bf f} $N_D=360$.
The same three regimes, pinned, fluctuating, and reordered, occur for all
values of $N_D$, but the location of the regimes shifts to different values
of $F_d$ as $N_D$ changes.
}
\end{figure}

{\bf Size effects on dynamical transitions:}
We measured dynamical reordering curves for samples with different numbers
of dislocations $N_D$.  The results in Fig. 2(a) are for a sample with 
$N_D=480$.  Figure S1 shows corresponding plots of
$P_{+-}$, $P_{--,++}$, and $B$ as a function of $F_d$ for samples with
$N_D=48$, 96, 144, 192, 240, and 360.  The three regimes, pinned, 
fluctuating, and reordered, illustrated in Fig. 2(a) also appear in the
smaller samples illustrated in Figure S1, but the transitions between
the regimes shift to lower values of $F_d$ as $N_D$ decreases.

\begin{figure}
\includegraphics[width = 3.5in]{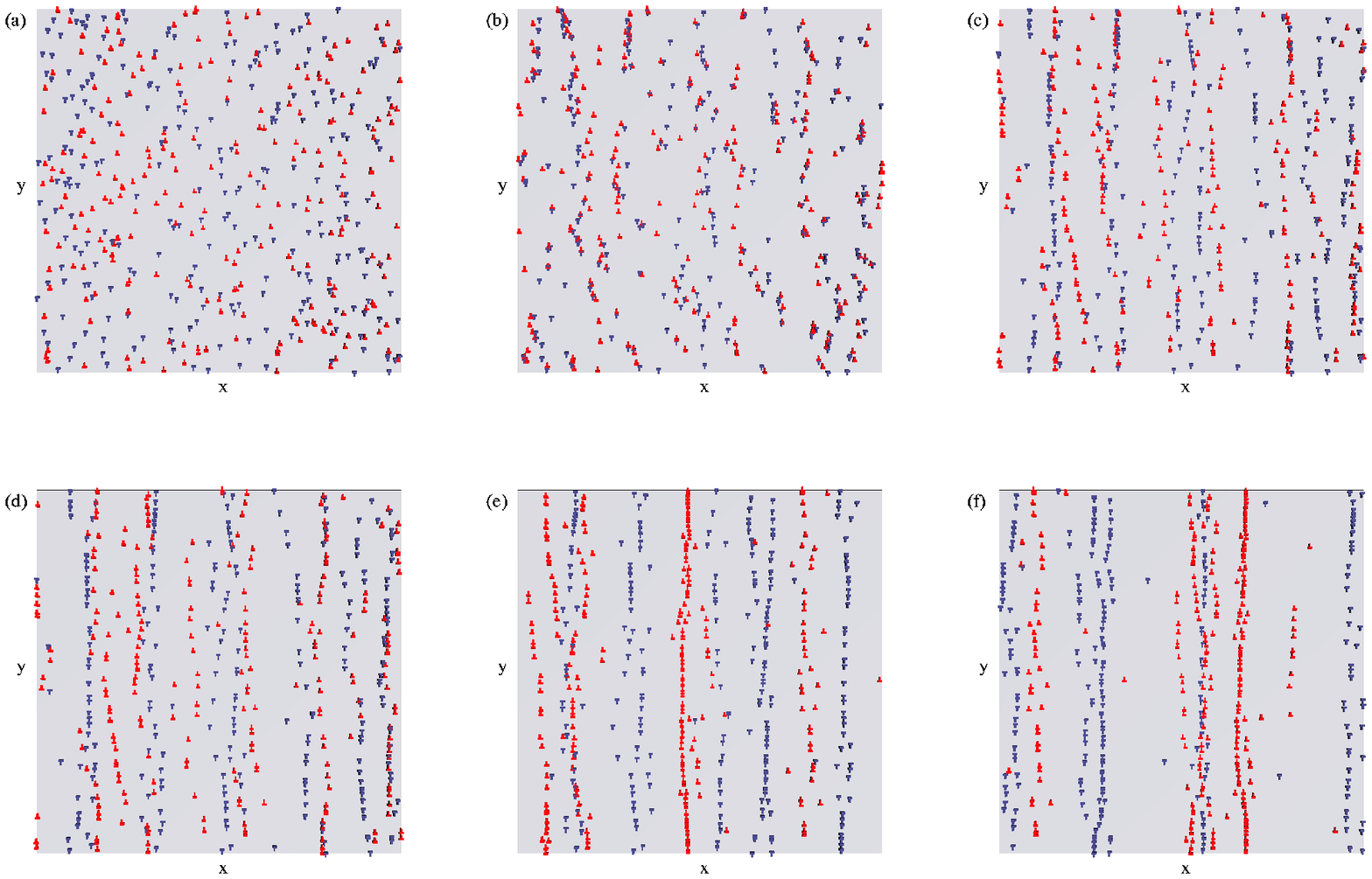}
\caption{
(Supplemental)
The red (blue) crosses are the locations of the dislocations with
positive (negative) Burgers vectors in a sample with $N_D=480$ that has
been instantaneously loaded with $F_d=10$.  The time progression of the
formation of multiple unipolar walls is illustrated.
{\bf a} Initial dislocation configuration.
{\bf b} After 1 time unit.
{\bf c} After 3 time units.
{\bf d} After 5 time units.
{\bf e} After 20 time units.
{\bf f} After 149 time units.
}
\end{figure}

{\bf Instantaneous loading:}
In a sample that is instantaneously subjected to a high load, we find a
transient disordered phase followed by the formation of multiple
lower density unipolar walls instead of the two unipolar walls shown in
Fig.~1(g).
For the sample with $N_D=480$, Figure S2 illustrates the time
progression of this process when we instantaneously increase $F_d$ to
$F_d=10$.



\begin{thebibliography}{99}

\bibitem{3bhattacharya}
S. Bhattacharya and M.J. Higgins, 
Phys. Rev. Lett. {\bf 70}, 2617
(1993).

\bibitem{4koshelev}
A.E. Koshelev and V.M. Vinokur, 
Phys. Rev. Lett. {\bf 73}, 3580
(1994).

\bibitem{5olson}
C.J. Olson, C. Reichhardt, and F. Nori, 
Phys. Rev. B {\bf 81}, 3757 
(1998).

\bibitem{8pardo}
F. Pardo, F. de la Cruz, P. Gammel, E. Bucher, and D.J. Bishop, 
Nature (London) {\bf 396}, 348
(1998).

\bibitem{9hellerqvist}
M.C. Hellerqvist, D. Ephron, W.R. White, M.R. Beasley, and A. Kapitulnik, 
Phys. Rev. Lett. {\bf 76}, 4022
(1996).

\bibitem{25troyanovski}
A.M. Troyanovski, J. Aarts, and P.H. Kes, 
Nature (London) {\bf 399}, 665
(1999).

\bibitem{7danneau}
R. Danneau, A. Ayari, D. Rideau, H. Requardt, J.E. Lorenzo, L. Ortega, 
P. Monceau, R. Currat, and G. Gr{\" u}bel, 
Phys. Rev. Lett. {\bf 89}, 106404 
(2002).

\bibitem{EC}
C. Reichhardt, C.J. Olson, N.J. Gr{\o}nbech-Jensen, and F. Nori, 
Phys. Rev. Lett. {\bf 86}, 4354
(2001).

\bibitem{26yaron}
U. Yaron {\it et al.}, 
Nature (London) {\bf 376}, 753
(1995)
Structural evidence for a two-step process in the depinning
of the superconducting flux-line lattice.
\emph{Nature (London)} 376:753-755.

\bibitem{6faleski}
M.C. Faleski, M.C. Marchetti, and A.A. Middleton, 
Phys. Rev. B {\bf 54}, 12427
(1996).

\bibitem{23balents}
L. Balents, M.C. Marchetti, and L. Radzihovsky, 
Phys. Rev. B {\bf 57}, 7705
(1998).

\bibitem{24ledoussal}
P. Le Doussal and T. Giamarchi, 
Phys. Rev. B {\bf 57}, 11356
(1998).

\bibitem{18moon}
K. Moon, R.T. Scalettar, and G.T. Zim{\' a}nyi, 
Phys. Rev. Lett. {\bf 77}, 2778
(1996).

\bibitem{moretti2004}
P. Moretti, M.-C. Miguel, M. Zaiser, and S. Zapperi,
Phys. Rev. B {\bf 69}, 214103
(2004).

\bibitem{laurson2010}
L. Laurson, M.-C. Miguel, and M.J. Alava, 
Phys. Rev. Lett. {\bf 105}, 015501
(2010).

\bibitem{bako2008}
B. Bak{\' o}, D. Weygand, M. Samaras, W. Hoffelner, and M. Zaiser, 
Phys. Rev. B {\bf 78}, 144104 
(2008).

\bibitem{2miguel}
M.-C. Miguel, A. Vespignani, S. Zapperi, J. Weiss, and J.R. Grasso,
Nature (London) {\bf 410}, 667
(2001).

\bibitem{12miguel}
M.-C. Miguel, A. Vespignani, M. Zaiser, and S. Zapperi, 
Phys. Rev. Lett. {\bf 89}, 165501 
(2002).

\bibitem{csikor2007}
F.F. Csikor, C. Motz, D. Weygand, M. Zaiser, and S. Zapperi, 
Science {\bf 318}, 251
(2007).

\bibitem{2008wang}
Z.Q. Wang, I.J. Beyerlein, and R. LeSar,
Phil. Mag. {\bf 88}, 1321 
(2008).

\bibitem{1tsekenis}
G. Tsekenis, N. Goldenfeld, and K.A. Dahmen, 
Phys. Rev. Lett. {\bf 106}, 105501
(2011).

\bibitem{1tsekenisA}
I. Groma, G. Gy{\" o}rgyi, and P.D. Ispanovity,
Phys. Rev. Lett. {\bf 108}, 269601 (2012);
G. Tsekenis, N. Goldenfeld, and K.A. Dahmen,
Phys. Rev. Lett. {\bf 108}, 269602 (2012).

\bibitem{16dimiduk}
D.M. Dimiduk, C. Woodward, R. LeSar, and M.D. Uchic, 
Science {\bf 312}, 1188 (2006).

\bibitem{14}
L. Laurson, and M.J. Alava,
Phys. Rev. E {\bf 74}, 066106 (2006).

\bibitem{20reichhardt}
C. Reichhardt and C.J. Olson Reichhardt, 
Phys. Rev. Lett. {\bf 103}, 168301
(2009).

\bibitem{21field}
S. Field, J. Witt, F. Nori, and X.S. Ling, 
Phys. Rev. Lett. {\bf 74}, 1206
(1995).

\bibitem{10marley}
A.C. Marley, M.J. Higgins, and S. Bhattacharya, 
Phys. Rev. Lett. {\bf 74}, 3029
(1995).

\bibitem{22olson}
C.J. Olson, C. Reichhardt, and F. Nori, 
Phys. Rev. B {\bf 56}, 6175 (1997).

\bibitem{27okuma}
S. Okuma, Y. Tsugawa, and A. Motohashi, 
Phys. Rev. B {\bf 83}, 012503 (2011).

\bibitem{2011ispanovity}
P.D. Isp{\' a}novity, I. Groma, G. Gy{\" o}rgyi, P. Szab{\' o},
and W. Hoffelner, 
Phys. Rev. Lett. {\bf 107}, 085506 (2011).

\bibitem{hirth}
J.P. Hirth and J. Lothe,
{\it Theory of Dislocations}
(Wiley, New York, 1982).

\bibitem{1987mughrabi}
H. Mughrabi, 
Mater. Sci. Eng. {\bf 85}, 15
(1987).

\bibitem{1996kuhlman}
D. Kuhlman-Wilsdorf, 
Scripta Mater. {\bf 34}, 641 
(1996).

\bibitem{1980kocks}
U.F. Kocks, T. Hasegawa, and R.O. Scattergood, 
Scripta Metal. {\bf 14}, 449
(1980).

\bibitem{1990stout}
M.G. Stout and A.D. Rollett, 
Metal. Trans. A {\bf 21}, 3201
(1990).

\bibitem{2007xue}
Q. Xue, I.J. Beyerlein, and D.J. Alexander, 
Acta Mater. {\bf 55}, 655
(2007).

\bibitem{2001huang}
X. Huang, A. Borrego, and W. Pantelon, 
Mater. Sci. Eng. A {\bf 319-321}, 237
(2001).

\bibitem{2000hughes}
D.A. Hughes and N. Hansen, 
Acta Mater. {\bf 48}, 2985
(2000).

\bibitem{2004dallatorre}
F. Dalla Torre, R. Lapovok, J. Sandlin, P.F. Thomson, C.H.J. Davies,
and E.V. Pereloma, 
Acta Mater. {\bf 52}, 4819
(2004).

\bibitem{2006dallatorre}
F.H. Dalla Torre, E.V. Pereloma, and C.H.J. Davies, 
Acta Mater. {\bf 54}, 1135
(2006).

\bibitem{Libook}
J.C.M. Li,
{\it Electron Microscopy and Strength of Crystals.}
(Interscience, New York, 1963).

\bibitem{Suppl}
See EPAPS document X.

\bibitem{17pertsinidis}
A. Pertsinidis and X.S. Ling, 
Phys. Rev. Lett. {\bf 100}, 028303 (2008).

\bibitem{2003kocks}
U.F. Kocks, and H. Mecking, 
Prog. Mater. Sci. {\bf 114}, 171
(2003).

\bibitem{1960li}
J.C.M. Li, 
Acta Metall. {\bf 8}, 296 (1960).

\end{thebibliography}
\end{document}